\documentclass[
  twocolumn,
  prl,nofootinbib,
  amsmath,
  amssymb,
  amsfont,
  superscriptaddress,
  longbibliography,
  floatfix]{revtex4-1}

\usepackage{bm,graphicx}
\graphicspath{{figs/}}

\begin{document}

\title{Disorder-induced rippled phases and multicriticality in free-standing graphene}

\author{D. R. Saykin}
\affiliation{Department of Physics, Stanford University, Stanford, CA 94305, USA}

\author{V. Yu. Kachorovskii}
\affiliation{Ioffe  Institute, Polytechnicheskaya 26, 194021, St.Petersburg, Russia}

\author{I. S. Burmistrov}
\affiliation{L. D. Landau Institute for Theoretical Physics, Semenova 1-a, 142432, Chernogolovka, Russia}
\affiliation{Laboratory for Condensed Matter Physics, National Research University Higher School of Economics, 101000 Moscow, Russia}

\begin{abstract}
One of the most exciting phenomena  observed in
crystalline disordered membranes, including a  suspended graphene, is  rippling, i.e. a formation
 of static flexural deformations. Despite an
 active research, it still remains
 unclear whether the rippled 
  phase exists in the thermodynamic
 limit, or  it is destroyed by thermal fluctuations. We demonstrate that a sufficiently strong
 short-range 
 disorder stabilizes ripples,
 whereas in the case of a weak disorder the thermal flexural
 fluctuations dominate  in the thermodynamic
 limit.
 The
 phase diagram of  the  disordered suspended graphene
 contains two  separatrices:
 the crumpling transition line dividing the
  flat and crumpled phases  and
   the rippling transition line demarking the rippled   and clean phases.
 At the intersection of the separatrices there is the unstable,
  multicritical point which splits up all four phases.
 Most remarkably, rippled and clean flat phases   are described by a {\it single}
 stable fixed point
 which belongs to  the rippling transition line.  Coexistence of two flat phases  in
 the single point is possible due to {\it non-analiticity} in corresponding renormalization group equations and reflects non-commutativity of limits of vanishing
  thermal and rippling fluctuations.

\end{abstract}


\maketitle

The study of critical elasticity of 2D crystalline membranes dates back to
the seminal paper by Nelson and Peliti \cite{Nelson1987}, where an idea of
crumpling transition (CT), i.e. the transition between flat and crumpled phases,
was put forward.  A more detailed analysis of
the CT and  anomalous elasticity of membranes has been  developed  in Refs.
\cite{Aronovitz1988,Paczuski1988,David1988,Guitter1988,Aronovitz1989,Guitter1989}.
The interest  to the field dramatically increased after
discovery of graphene
\cite{Geim,Geim1,Kim}.
A suspended graphene (for a review, see Refs.
\cite{geim07,novoselov07,graphene-review,review-DasSarma,review-Kotov,book-Katsnelson,
book-Wolf,book-Roche}) 
provides
an excellent opportunity not only to experimentally
verify the existing theoretical predictions for
two-dimensional (2D) crystalline membranes but to challenge
 the theory by new unexpected experimental data.
The  underlying  physics of CT is determined by the
thermal out-of-plane fluctuations, so-called  flexural phonons (FP). On the one hand,  FP tend to  crumple
the membrane. On the other hand, the long--range interactions between FP, i.e. anharmonic effects, ``iron'' the membrane
and stabilize the flat phase. As a result of such competition,
flat and crumpled phases can exist in a clean crystalline membrane. 

Along with FP, there can subsist the  static, frozen deformations, the so-called ripples, caused by imperfection of the crystal lattice. Such deformations act similarly to FP
and also tend to crumple  the membrane
as was predicted long time ago
\cite{Morse:1992,Nelson_1991,Radzihovsky:1991,Morse:1992b,Bensimon_1992,Bensimon_1992}. However the physics of disordered membranes
with a non-trivial interplay of ripples and thermal fluctuations
 is  much less understood as compared to the clean case. In particular,
it is not even  fully resolved how many phases exist in such membranes.

The competition between thermal fluctuations and ripples is
of crucial importance  for free-standing
graphene. 
Indeed, the effect of the thermal fluctuations is controlled
 by the ratio of temperature $T$ and the bending
rigidity $\varkappa_0$. In a clean 2D membrane the CT occurs at $T / \varkappa_0 \sim 1$.
In graphene, $\varkappa_0\sim 1$ eV, so that the thermal fluctuations alone are not enough to crumple it.  
At the same time, recent numerical simulations of disordered graphene clearly show  the CT
\cite{Giordanelli2016}.
Additional evidence for importance of disorder in graphene is provided by
recent experimental measurements of anomalous Hooke's law (AHL)
\cite{Nicholl2015,Nicholl2017}.
Measured scaling exponent was substantially
different from the one known from  numerical simulations  for the clean
case \cite{Los2016}. These experimental and numerical results  imply existence of the rippled
phase
with properties distinct from the clean one.

\begin{figure}[b]
\centerline{\includegraphics[width=0.97\columnwidth]{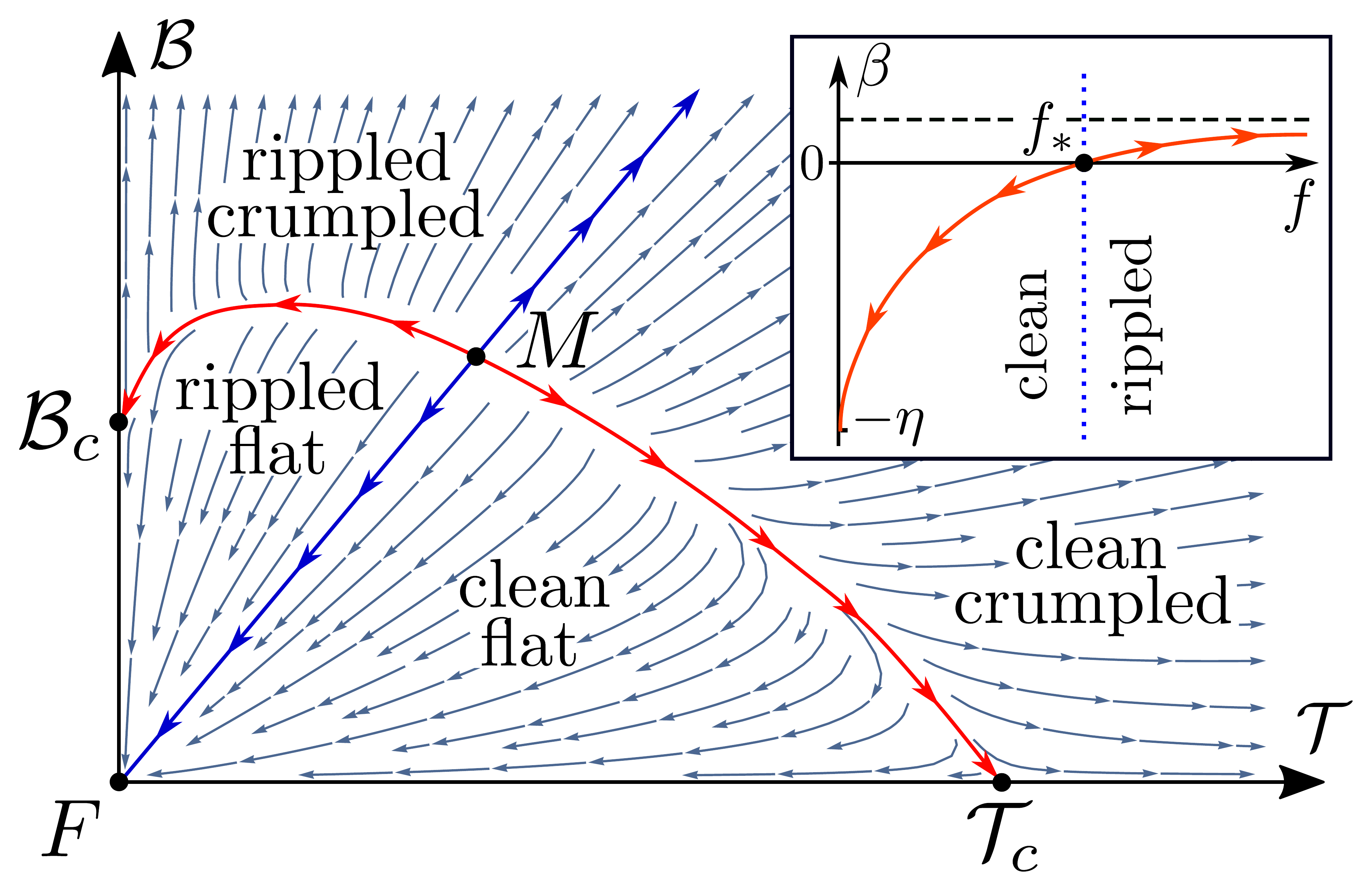}}
\caption{(Color online) Sketch of the phase diagram in  plane of the rescaled disorder strength
$\mathcal{B}=b/\xi^2$ and  the amplitude of thermal fluctuations $\mathcal{T}= T/(\varkappa \xi^2).$
The solid red curve corresponds to the crumpling transition. The solid blue line
is the rippling transition line, separating disordered and clean phases.
Fully unstable (multicritical) point is marked by $M$. Two  flat phases (clean and disordered) coexists in  the stable singular fixed  point $F$ reflecting non-commutativity of limits of vanishing thermal and rippling fluctuations. The arrows mark direction of RG flow towards the infrared. Inset: The sketch of a one-parameter RG flow for $f=\mathcal{B}/\mathcal{T}$.
  }
\label{Figure1}
\end{figure}

Previous theoretical studies of disordered 2D  membranes
\cite{Morse:1992b,Bensimon_1992} predict  existence of the rippled flat phase
at exactly zero temperature, $T=0$, which is unstable with
respect to the thermal fluctuations (similar conclusion has been obtained
for disordered $D=4-\epsilon$  dimensional membrane
\cite{Morse:1992,Nelson_1991,Radzihovsky:1991}).
This conclusion implies absence of the stable rippled phase,
and, at first glance, contradicts to observations of
Refs.~\cite{Giordanelli2016,Nicholl2015,Nicholl2017,Los2016}.
Recently, the CT in disordered suspended graphene (DSG) was
addressed in Ref.~\cite{Gornyi:2015a}.
It was shown that disorder can crumple a membrane in agreement with Ref.~\cite{Giordanelli2016}.
It was also found~\cite{Gornyi:2015a} that  instability of  the rippled phase  predicted  in
Refs.~\cite{Morse:1992b,Bensimon_1992} develops
logarithmically slow, i.e. the marginal $T=0$ rippled phase controls
elastic properties of DSG for $T\neq 0$ in a wide interval of length scales (see also Ref. \cite{Doussal2018}).
This   marginal behavior  can manifest itself in  experiments on  AHL in graphene
\cite{Nicholl2015,Nicholl2017} as was  demonstrated in Ref.~\cite{Gornyi2017}.
However, the rippled phase should  not ``survive'' in the
thermodynamic limit even for the case of very strong disorder.
Alternatively, observations of Refs.~\cite{Giordanelli2016,Nicholl2015,Nicholl2017} can indicate the
existence of a stable rippled flat phase at finite temperature. Therefore, a
phase diagram of a DSG, when ripples and thermal fluctuations are competed, remains to  be still
established.

In this Letter, we report the phase diagram of a 2D crystalline membrane
with short-ranged curvature disorder  (see Fig. \ref{Figure1}).
Our main results are as follows.
\begin{itemize}
\item
There are four distinct phases:
clean/rippled flat and
clean/rippled crumpled ones. There is a fully unstable,  {\it multicritical }  fixed point
marked by  $M$  that  splits up all four phases. 
\vspace{-0.2cm}
\item
There is   a  stable  fixed point $F$ corresponding simultaneously
to clean and rippled flat phases. Coexistence of two flat phases  in
 the single fixed point reflects non-commutativity of limits of vanishing
  thermal and rippling fluctuations and is possible
  due to {\it singularity} in corresponding renormalization group (RG)  equations, cf. Eq. \eqref{eq:RGtb}. 
\vspace{-0.2cm}
\item
There are two  separatrices: one corresponding to the CT (red solid curve) and the other separating clean and rippled  phases (blue solid curve).
\vspace{-0.2cm}
\end{itemize}

To obtain this results,  we performed a standard $1/d_c$ expansion \cite{David1988}
up to the second order, where $d_c$ is the number of FP.
As was recently demonstrated
\cite{Burmistrov2018a,Burmistrov2018b,Burmistrov2019a}, the second order
diagrams  contain  ones that are not accounted  by the so-called
Self-Consistent Screening Approximation (SCSA)
~\cite{Doussal1992} which is frequently discussed as an efficient
approximate scheme
\cite{Gazit2009,Doussal2018}.  Our results represent first
rigorous treatment of  anharmonicity in disordered  membranes  within
second-order in $1/d_{\rm c}$ expansion, which is not accounted for neither by SCSA, nor
by other approximative schemes such as  non-perturbative RG approach   \cite{Kownacki2009,Braghin2010,Coquand:2018}.
We demonstrate  that  the  finite
temperature instability of the rippled phase was an artefact of
the first order approximation in $1/d_{\rm c}.$
Our key  technical finding is that   the terms of higher order in $1/d_{\rm c}$
stabilize the rippled marginal phase  and lead to the appearance
of the rippling transition (RT) line shown in Fig.~\ref{Figure1}.

\textit{Disorder.} ---There are many ways to introduce a disorder experimentally:
by bombarding
 graphene
 with heavy atoms \cite{Yeo2018},   by  fluorination \cite{Daukiya2016},
 or by
 creating macroscopical  defects, e.g. artificial holes  \cite{Yllanes2017}.
Theoretically, one classifies disorder with respect to the reflection
symmetry related the two opposite sides of a membrane.
An example of disorder which preserves the reflection
symmetry is the so-called metric or in-plane disorder. It can arise due to
the fluctuations in concentration of impurity atoms.
Such short-ranged disorder is irrelevant  at $T\neq 0$ in the thermodynamic limit,
i.e.
the clean flat phase is stable against an in-plane disorder
\cite{Nelson_1991,Radzihovsky:1991,Gornyi:2015a} (for discussion
of special case $T= 0$, see Ref.~\cite{Radzihovsky_1992}).
Therefore, we do not consider metric disorder here.
Instead, we consider  a random curvature disorder
proposed in
Refs. \cite{Morse:1992,Morse:1992b,Bensimon_1992}, which
breaks the reflection symmetry. Such disorder naturally
arises
if impurity atoms are situated on one side of a membrane. 

\textit{Energy functional.} --- 
Membrane's configuration is parameterized with vector $\bm{r}(\bm{x}) \in \mathbb{R}^d$, $\bm{x} \in \mathbb{R}^D$ where $D=d-d_c$. We introduce stretching factor $\xi_0$, which characterizes the projective area of a membrane, $\xi_0^2L^2$, and use vectors $\bm{u}(\bm{x}) \in \mathbb{R}^D$ and $\bm{h}(\bm{x}) \in \mathbb{R}^{d_c}$ to describe in-plane and out-of-plane displacements:
$\bm{r} = \xi_0 \bm{x}+\bm{u}+\bm{h}$. Although, in the case of graphene $D=2$ and the number of FP is one, we consider
$d_c$ as an arbitrary parameter which allows us to develop controllable perturbation theory in $1/d_c$ \cite{David1988}.
The energy of crystalline membrane consist of bending and elastic contributions \cite{Nelson1987,Morse:1992,Morse:1992b,Bensimon_1992}
\begin{gather}
\mathcal{F} = \! \int \!d^2 \bm{x} \left [  \frac{\varkappa_{0}}{2}    (\Delta \bm{h} - \bm{\beta})^2
    + \mu_0 u_{\alpha\beta}^{2}+\frac{\lambda_0}{2} u_{\alpha\alpha}^{2} \right ].
\label{eq:action:0}
\end{gather}
Here $\mu_0$ and $\lambda_0$ stand for the Lam\'e coefficients. 
The last two
terms in the right hand side (r.h.s.) of Eq. \eqref{eq:action:0}  describe in-plane elastic
energy with the strain tensor $u_{\alpha\beta} = (\partial_\alpha u_\beta + \partial_\beta u_\alpha + \partial_\alpha \bm{h} \partial_\beta \bm{h})/2$. Quenched random curvature is added via a zero-mean Gaussian random vector $\bm{\beta}$ \cite{Morse:1992,Morse:1992b,Bensimon_1992}. Strength of disorder is controlled by a variance $b_0$: 
$\overline{\beta_j(\bm{x})\beta_k(\bm{x}^\prime)}= b_{0} \,
\delta_{jk} \delta(\bm{x}-\bm{x}^\prime)$, $j,k=1,\dots,d_c$.
%

\begin{figure*}[t]
\centerline{
	\includegraphics[height=0.075\textheight]{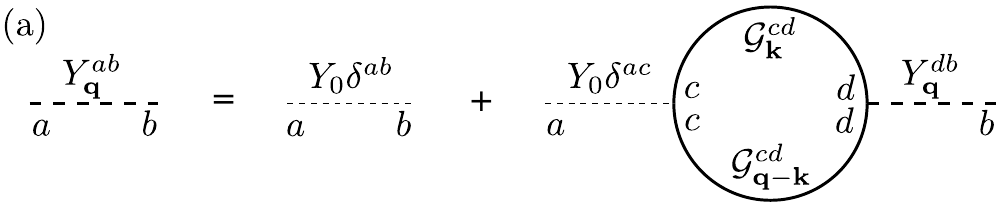} \hfil
	\includegraphics[height=0.075\textheight]{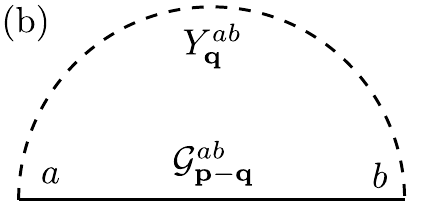}
}
\caption{(a) The equation for the screened interaction. The solid line represents the bare Green's function $\hat{\mathcal{G}}$. The thin dashed line denotes the bare interaction proportional to $Y_0$. The dashed line stands for the screened interaction $Y_q$. (b) The self-energy correction of the first order in $1/d_c$.}
\label{Figure-D2}
\end{figure*}

\textit{Anomalous elasticity.} ---
Although  the CT cannot be observed in a clean graphene,
the thermal fluctuations around the flat phase
lead to highly
non-trivial anomalous elastic properties in membranes of size
above  the so-called Ginzburg length $L _ *\sim \varkappa_0/\sqrt{Y_0 T}
$ \cite{Aronovitz1989}.
Here $Y_0= \frac{4\mu_0(\mu_0+\lambda_0)}{(2\mu_0+\lambda_0)}$  denotes the bare (ultraviolet) value of the 2D Young's modulus.
Due to an anharmonic coupling between the in-plane and out-of plane elastic
modes the bending  rigidity increases in a power law manner for $L\gg L_*$  with a
certain critical exponent $\eta$: $\varkappa \propto \varkappa_0 (L / L _ *) ^\eta
$.
For graphene $L_*\sim 1\div 10$ nm, so that realistic flakes of graphene
 are in the regime of anomalous elasticity and show
a number of highly non-trivial phenomena already verified
experimentally, such as  mentioned above AHL \cite
{Guitter1988,Aronovitz1989,Lopez2015,Nicholl2015,Los2016,Nicholl2017,Lopez2017,Gornyi2017},
 negative thermal expansion coefficient
\cite{Zakharchenko2009,Bao2009,Yoom2011,Andres2012,Silva2014,Michel2015,Burmistrov2016},
power-law scaling of the phonon-limited conductivity \cite
{Bolotin2008,Castro2010,Gornyi2012}, etc.

\textit{Crumpling transition.} ---
The scaling equation describing dependence of the stretching factor on a membrane size $L$ reads \cite{Gornyi:2015a}
\begin{equation}
\frac{d\xi^2}{d\ln L} =
- \frac{d_c}{4\pi}\left ( {\cal T +B} \right ) \xi^2
,\qquad  \xi(L_*) =\xi_0,
\label{eq:xi:RG}
\end{equation}
where $\mathcal{T}=T/\varkappa \xi^2$ and  $\mathcal{B}=b/\xi^2$ are rescaled amplitudes of the thermal and disorder--induced
fluctuations. The CT occurs when $\xi$ turns into zero at a finite length scale, while in the
flat phase $\xi(L\to \infty)>0$.
Both the thermal and rippling fluctuations tend to  crumple
the membrane.
In the clean case, $\mathcal{B}=0$, the power law dependence of $\cal{T}$ on $L$
results in the crumpling transition at a very high temperature
$T_c=4\pi \eta \varkappa_0/d_c$, which is unreachable for graphene.

In a disordered membrane scaling of ${\cal T}$ and ${\cal B}$ is more intricate than a power law.
There is a CT  curve in the plane (${\cal T}, {\cal B}$) which was found in Ref.~\cite{Gornyi:2015a}
by using first order  expansion over $1/d_{\rm c}$. Next, we demonstrate that higher order terms in $1/d_{\rm c}$ lead to appearance of  the rippling transition
line (blue line in Fig.~\ref{Figure1}).

Replicating  fields $\bm{u}$ and $\bm{h}$, integrating over $\bm{u}$  and performing averaging
of the replicated partition function over disorder, we obtain the
effective free energy \cite{Gornyi:2015a}.
\begin{gather}
\mathcal{F}_{\rm dis}= \!\sum_{a,b=1}^N  \int\! \frac{d^2\bm{k}}{(2\pi)^2} \frac{\hat \varkappa_{ab}\, \bm{k}^4}{2} \bigl (\bm{h}_{\bm{k}}^{(a)}\bm{h}^{(b)}_{-\bm{k}}\bigr ) +\frac{Y_{0}}{8} \sum_{a=1}^N\int\! \frac{d^2 {\bm q}}{(2\pi)^2}
\notag \\
\times
\left |\int \frac{d^2 {\bm k}}{(2\pi)^2} \frac{[\bm{k}\times \bm{q}]^2}{q^2}
    \bigl (\bm{h}^{(a)}_{\bm{k}+\bm{q}} \bm{h}^{(a)}_{-\bm{k}}\bigr )\right |^2 ,
\label{eq:action:2}
\end{gather}
where $\hat{\varkappa}_{ab} =
\varkappa_{0} \bigl [ \delta_{ab} - f_0 J_{ab} \bigr ]$ with matrix $\hat{J}$ having all entries equal to one, indices $a,b=1,\dots,N$ enumerate replicas,  and $f_0= b_0\varkappa_0/T$.

\begin{figure}[t]
\centerline{
	\includegraphics[height=0.075\textheight]{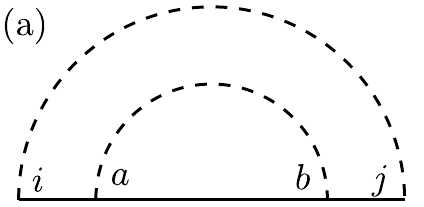}\hfil
	\includegraphics[height=0.075\textheight]{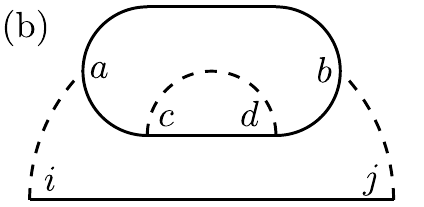}
}\vspace{0.01\textheight}
\centerline{
	\includegraphics[height=0.075\textheight]{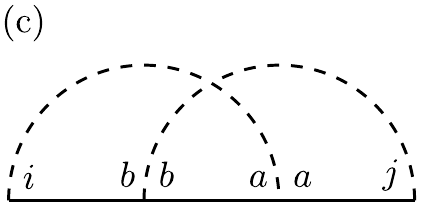}\hfil \includegraphics[height=0.075\textheight]{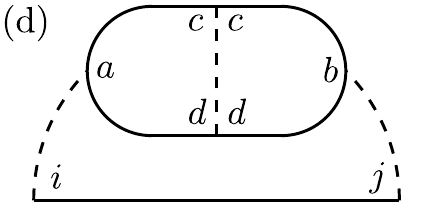}
}\vspace{0.01\textheight}
\centerline{
	\includegraphics[height=0.075\textheight]{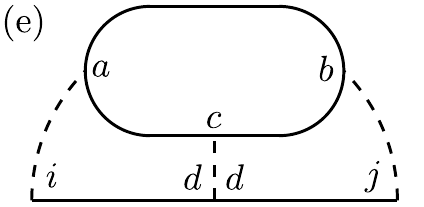}\hfil 
 	\includegraphics[height=0.075\textheight]{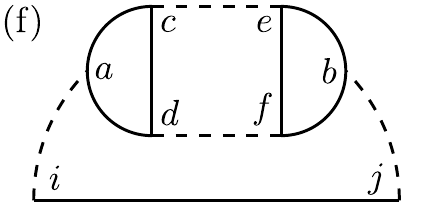}
 }
\caption{Diagrams for the self-energy corrections in the second order in $1/d_c$. Diagrams (a) and (b) are included in SCSA. Diagrams (c)-(f) are not taken into account by SCSA.}
\label{Figure-D3}
\end{figure}

Anharmonicity of FP result in a renormalization of the parameters of $\mathcal{F}_{\rm dis}$. The necessary information can
be extracted from the exact two-point Green's function  $ \langle
h_i^{(a)}(\bm{k}) h_j^{(b)}(-\bm{k})\rangle \equiv
\hat{\bm{\mathcal{G}}}_{ab}(k) \delta_{ij}$ 
where the average is with
respect to the free energy \eqref{eq:action:2}. The quadratic part of
$\mathcal{F}_{\rm dis}$ determines the bare Green's function
$\hat{\mathcal{G}}_{ab}(k) = T(\delta_{ab}+f_0 J_{ab})/(\varkappa_0 k^4)$. At first,
the screening of  the interaction between flexural phonons should be taken
into account via RPA-type resummation (see Fig. \ref{Figure-D2}). The
screened interaction becomes independent of $Y_0$ for
$q<\sqrt{d_c (1+2f_0)}/L_*$ \cite{Gornyi:2015a} and behaves as
$q^2/d_c$ as $q\to 0$.
Using this screened interaction we can construct
the regular perturbation theory in $1/d_c$ for the self-energy
$\hat\Sigma$ (see diagrams in Figs. \ref{Figure-D2} and \ref{Figure-D3})
which relates the exact and bare Green's functions:
$\hat{\bm{\mathcal{G}}}^{-1} = \hat{\mathcal{G}}^{-1}-\hat \Sigma$. 
The
perturbation theory for $\hat\Sigma(k)$ has infrared logarithmic
divergences as $k\to 0$. They can be used to extract the 
(RG) behavior of the theory.

\textit{RG flow.} --- The corresponding  RG equations can be written in the following form.
\begin{equation}
\frac{d\ln \varkappa}{d\ln L}  = \eta_\varkappa(f)
, \,\,  \frac{d \ln b}{d\ln L}  = -\eta_b(f) ,\,\, \frac{d \ln f}{d\ln L}  =  \beta(f) ,
\label{eq:RG:eqs}
\end{equation}
where $f= b\varkappa/T=\mathcal{B}/\mathcal{T}$ and $\eta_b(f)= \eta_\varkappa(f)-\beta(f)$.
We emphasize that RG equation for $f$ decouples,
while $\varkappa$  and $b$  are slave variables.
The RG functions can be expanded
as $\eta_{\varkappa/b}={\eta_{\varkappa/b}^{(1)}}/{d_c}
 +
{\eta_{\varkappa/b}^{(2)}}/{d_c^2} + \dots$ and $\beta={\beta^{(1)}}/{d_c}
 +
{\beta^{(2)}}/{d_c^2} + \dots$. 
The first order coefficients are $\beta_1= -2(1+3f)/(1+2f)^2$ and
$\eta_\varkappa^{(1)}= 2(1+3f+f^2)/(1+2f)^2$ \cite{Morse:1992b,Gornyi:2015a}.
Our explicit calculations in the second order in $1/d_c$ yield \cite{See Supplemental Material}
\begin{align}
\beta^{(2)}=& -\Bigl [73 + 803 f + 3667 f^2 + 8517
f^3 + 9278 f^4  \notag \\ +& 3420 f^5
+ 186 f^6 -68\zeta(3) \bigl (1 + 11 f + 49 f^2 \notag \\
+ &
111 f^3 + 128 f^4
+  58 f^5 + 6 f^6\bigr )\Bigr]\Bigl /\bigl [27(1+2f)^6\bigr ] ,
\notag \\
\eta_\varkappa^{(2)}= &
\Bigl [73 + 803 f + 3550 f^2 + 7743 f^3 + 7995 f^4 +3046 f^5   \notag \\ +&  265 f^6 -68\zeta(3) \bigl (1 + 11 f + 49 f^2+111 f^3+129 f^4  \notag \\
+ & 64 f^5 + 9 f^6\bigr )\Bigr]\Bigl /\bigl [27(1+2f)^6\bigr ]  .
 \end{align}
The functions $\beta^{(2)}$ and $\eta_\varkappa^{(2)}$ have finite limit at $f\to \infty$.
The RG flow for $f$ has three fixed points: $0$, $\infty$, and $f_*
\approx 8.5 \, d_c $ (see the inset to Fig. \ref{Figure1})
\cite{footnote}.
At the fixed point $f=0$
which is {\it stable} in the infrared
the bending rigidity and disorder variance acquires the power
law scaling, $\varkappa \sim L^{\eta}$,  $b\sim L^{-\eta^\prime}$ where \cite{Burmistrov2019a}
\begin{equation}
\eta = \eta^\prime/2 = \eta_\varkappa(0)\approx2/d_c - [68\zeta(3)-73]/(27 d_c^2) .
\label{eq:scaling:0}
\end{equation}
The other infrared {\it stable} fixed point is located at $f=\infty$.
We note that within the first order in $1/d_c$ this fixed point is marginally unstable \cite{Morse:1992b}. 
At $f=\infty$ the bending rigidity and the
disorder variance has also power-law scaling with momentum, $\varkappa \sim L^{\eta_\infty}$,  $b\sim L^{-\eta^\prime_\infty}$ where
\begin{equation}
 \eta_\infty  \approx \frac{1}{2d_c} - \frac{612 \zeta(3)-265}{1728 d_c^2},
 \,
 \eta^\prime_\infty
 \approx \eta_\infty -
\frac{68\zeta(3)-31}{228 d_c^2} .
\label{eq:scaling:1}
\end{equation}
The fixed point at $f=f_*$ is {\it unstable} in the infrared and
is characterized by the exponent of divergent
correlation length $\nu = 1/[f d \beta/df]|_{f=f_*} = 288
d_c^2/[68\zeta(3)-31]$ \cite{Comment-Mohana}.

From Eqs. \eqref{eq:xi:RG} and \eqref{eq:RG:eqs} we find the RG equations governing the flow of parameters $\mathcal{T}$ and $\mathcal{B}$.
\begin{equation}
\begin{split}
\frac{d \ln {\cal T}}{d\ln L} & =\frac{d_c}{4\pi} ({\cal B}+ {\cal T})-\eta_\varkappa({\cal B}/{\cal T}),
 \\
\frac{d \ln {\cal B}}{d\ln L} & =\frac{d_c}{4\pi} ({\cal B}+{\cal T})-\eta_b({\cal B}/{\cal T}).
\end{split}
\label{eq:RGtb}
\end{equation}
The corresponding flow diagram is shown in Fig.~\ref{Figure1}.
There is the {\it unstable} fixed point at ${\cal T}_c=4\pi \eta/d_c$ and
${\cal B}=0$, corresponding to the CT due to thermal
fluctuations in the absence of disorder. This fixed point controls
the transition between the clean flat phase and the clean crumpled
phase.
The {\it unstable} fixed point at $ {\cal T}   $ and ${\cal B}_{c}=4\pi
\eta^\prime_\infty/d_c$ corresponds to the  disorder-driven CT
 \cite{Gornyi:2015a}.
This fixed point controls transition between flat and crumpled rippled phases.
Remarkably,
both the clean flat phase ($f=0$) and the rippled flat phase ($f=\infty$)
are described by the {\it single singular infrared stable} fixed point $F$ at
${\cal T}={\cal B}=0.$
The  singularity at $ {\cal T} \to 0$ and ${\cal B} \to 0$  is clearly seen if one uses
 in Eq.~\eqref{eq:RGtb}, the  explicit expressions for the functions $\eta_{\varkappa} ({\cal B}/{\cal T})$
 and $\eta_b({\cal B}/{\cal T})$. 
 Eqs. \eqref{eq:RGtb} admit the {\it
multicritical} fixed point $M$ at ${\cal T}_*= 4\pi \eta_\varkappa(f_*)/[d_c(1+f_*)]$ and
${\cal B}_*=f_* {\cal T}_*$ (see Fig.~\ref{Figure1}). This multifractal fixed point has two unstable
directions: along the CT curve, which demarks  flat and crumpled phases,  and  along the RT line,
${\cal B}=f_* {\cal T},$ which splits up clean and rippled phases and  connects $M$ and $F$.
The scaling along these two separatrices  are  controlled by the critical
exponents $\nu$ and  $1/\eta_\varkappa(f_*),$ respectively. 
We emphasise the striking resemblance of our RG flow diagram with the
one for the random bond Ising model \cite{Nishimori}. The RT
line corresponds to the so-called Nishimori line \cite{LeDoussal1988}.

\textit{Discussion and conclusion.} ---  Our key result  is the
demonstration  of the ripples stabilization by sufficiently strong
disorder. More precisely, two  transitions occur with increasing the
disorder at fixed other parameters. For ${\cal T}< {\cal T}_*$ (see
Fig.~\ref{Figure1}), the first transition corresponds to the stabilization
of ripples, while the second one  is the CT.  On the contrary, for ${\cal
T}> {\cal T}_*,$ the CT happens before the stabilization of ripples. Our
phase diagram suggests also  a possibility of the RT with
decreasing temperature at the fixed disorder. As a very interesting
subject for further research, we expect that the phase diagram is even
reacher in the case of long-ranged  disorder \cite{LD&R1993}.

In the course of derivation of  RG Eqs. \eqref{eq:RGtb} we neglected the term $\partial_\alpha \bm{u}\partial_\beta \bm{u}$ in the expression for the strain tensor $u_{\alpha\beta}$. It can be shown \cite{Future} that this approximation is justified for $\mathcal{B}, \mathcal{T}\ll 1$. The terms $\partial_\alpha \bm{u}\partial_\beta \bm{u}$ provide additional contributions to Eqs. \eqref{eq:RGtb} that are of higher order in powers of $\mathcal{B}$ and $\mathcal{T}$. These terms do not affect the properties of the fixed point F but can result in corrections of higher order in $1/d_c$ to the position of the CT line as well as to the critical exponents governing scaling behavior at the fixed points $\mathcal{B}_c$, $\mathcal{T}_c$, and M \cite{Future}.

The relevance of our theory for realistic graphene membranes is supported by
 the numerical simulations \cite{Giordanelli2016} where the CT with increase of disorder was clearly seen and the fractal dimension  of  the crumpled membrane was reported.
A detailed comparison of our theory with Ref. \cite{Giordanelli2016}
 is however  not possible due to a lack of simulations at various temperatures.

To conclude, we predict existence of  two disorder-dominated  rippled phase
(flat and crumpled)
 in a disordered crystalline membrane with a short-range disorder.  By using fully
controlled  standard $1/d_{c}$ expansion, we
derive coupled RG equations for
the bending  rigidity and disorder strength and
establish the phase diagram of a generic crystalline membrane
(see Fig.~\ref{Figure1}).
We demonstrate existence of the
multicritical point (M),  the singular stable point (F), where the rippled flat
and clean flat phases coexist, and the rippling transition line connecting these two fixed points.

\begin{acknowledgements}
We thank I. Gornyi, I. Gruzberg, A. Mirlin for useful discussions. The
work was funded in part by the Alexander von Humboldt Foundation, by
Russian Ministry of Science and Higher Educations, the Basic Research
Program of HSE, and by the Russian Foundation for Basic Research (grant No. 20-52-12019) -- Deutsche Forschungsgemeinschaft (grant No. SCHM 1031/12-1) cooperation. The authors are grateful Institute for Theory of Condensed Matter of Karlsruhe Institute of Technology for hospitality.
\end{acknowledgements}


\begin{thebibliography} {128}

\bibitem{Nelson1987} D. Nelson and L. Peliti, {\it Fluctuations in membranes with crystalline and hexatic order}, J. Phys., {\bf 48}, 1085
(1987).

 \bibitem{Aronovitz1988} J. A. Aronovitz and T. C. Lubensky, {\it Fluctuations of solid membranes}, Phys. Rev. Lett.
{\bf 60}, 2634 (1988).

 \bibitem {Paczuski1988} M. Paczuski, M. Kardar, and D. R. Nelson, {\it Landau theory of the crumpling transition}, Phys. Rev.
Lett.{\bf  60}, 2638 (1988).

\bibitem {David1988} F. David and E. Guitter, Europhys. Lett. (EPL), {\it Crumpling transition in elastic membranes: Renormalization group treatment},
    {\bf 5}, 709 (1988).

\bibitem{Guitter1988} E. Guitter, F. David, S. Leibler, and L. Peliti, {\it Crumpling and buckling transitions in polymerized membranes}, Phys. Rev. Lett. {\bf 61}, 2949 (1988).

\bibitem {Aronovitz1989} J. Aronovitz, L. Golubovic, and T. C. Lubensky, {\it Fluctuations and lower critical dimensions of crystalline membranes}, J. Phys. {\bf 50}, 609 (1989).

\bibitem {Guitter1989} E. Guitter, F. David, S. Leibler, and L. Peliti, {\it Thermodynamical behavior of polymerized membranes}, J. Phys. {\bf 50}, 1787 (1989).

\bibitem{Geim} K. S. Novoselov, A. K. Geim, S. V. Morozov, D. Jiang, Y. Zhang, S. V. Dubonos, I. V. Grigorieva, and A. A. Firsov, {\it Electric field effect in atomically thin carbon films}, Science {\bf 306}, 666 (2004).

\bibitem{Geim1} K. S. Novoselov, A. K. Geim, S. V. Morozov, D. Jiang, M. I. Katsnelson, I. V. Grigorieva, S. V. Dubonos, and A. A. Firsov, {\it Two-dimensional gas of massless Dirac fermions in graphene}, Nature \textbf{438}, 197 (2005).
    
\bibitem{Kim} Y. Zhang, Y.-W. Tan, H. L. Stormer, and P. Kim, {\it Experimental observation of the quantum Hall effect and Berry's phase in graphene}, Nature
    \textbf{438}, 201 (2005).
    
\bibitem{geim07} A. K. Geim and K. S. Novoselov, {\it The rise of graphene}, Nature Materials {\bf 6}, 183 (2007).

\bibitem{novoselov07} K. S. Novoselov, Z. Jiang, Y. Zhang, S. V. Morozov, H. L. Stormer, U. Zeitler, J. C. Maan, G. S. Boebinger, P. Kim, and A. K. Geim, {\it Room-temperature quantum Hall effect in Graphene}, Science {\bf 315}, 1379 (2007).

\bibitem{graphene-review} A. H. Castro Neto, F. Guinea, N. M. R. Peres, K. S. Novoselov, and A. K. Geim, {\it The electronic properties of graphene}, Rev. Mod. Phys.  {\bf 81}, 109 (2009).

\bibitem{review-DasSarma} S. Das Sarma, S. Adam, E. H. Hwang, and
    E. Rossi, {\it Electronic transport in two-dimensional graphene}, Rev. Mod. Phys. {\bf 83}, 407 (2011).
    
\bibitem{review-Kotov} V. N. Kotov, B. Uchoa, V. M. Pereira, F. Guinea, and A. H. Castro Neto, {\it Electron-electron interactions in graphene: Current status and perspectives}, Rev. Mod. Phys. {\bf 84}, 1067 (2012).

\bibitem{book-Katsnelson} M. I. Katsnelson, {\it Graphene: Carbon in Two
    Dimensions}, Cambridge University Press  (2012).
    
\bibitem{book-Wolf} E. L. Wolf, {\it  Graphene: A New Paradigm in
    Condensed Matter and Device Physics}, Oxford University Press (2014).

\bibitem{book-Roche} L. E. F. Foa Torres, S. Roche, J.-C. Charlier,  {\it
Introduction to Graphene-Based Nanomaterials From Electronic Structure to
Quantum Transport},  Cambridge  University Press (2014).

\bibitem {Morse:1992} D. C. Morse, T. C. Lubensky, and G. S. Grest, {\it Quenched disorder in tethered membranes}, Phys.
    Rev. A {\bf 45}, R2151 (1992).

 \bibitem {Nelson_1991} D. R. Nelson and L. Radzihovsky, {\it Polymerized membranes with quenched random internal disorder}, Europhys. Letters
(EPL) {\bf 16}, 79 (1991).

\bibitem {Radzihovsky:1991}  L. Radzihovsky and D. R. Nelson, {\it Statistical mechanics of randomly polymerized membranes}, Phys. Rev. A {\bf 44}, 3525 (1991).

\bibitem {Morse:1992b} D. C. Morse and T. C. Lubensky, {\it Curvature disorder in tethered membranes: A new flat phase at T=0}, Phys. Rev. A {\bf 46}, 1751 (1992).

\bibitem {Bensimon_1992} D. Bensimon, D. Mukamel, and L. Peliti, {\it Quenched curvature disorder in polymerized membranes}, Europhys. Letters (EPL) {\bf 18}, 269 (1992).

\bibitem{Giordanelli2016} I. Giordanelli, M. Mendoza, J. S. Andrade Jr., M. A. F. Gomes, and H. J. Herrmann, {\it Crumpling damaged graphene}, Scientific Reports, {\bf 6}, 25891 (2016).

\bibitem{Nicholl2015} R. J. T. Nicholl, H. J. Conley, N. V. Lavrik, I. Vlassiouk, Y. S.
    Puzyrev, V. P. Sreenivas, S. T. Pantelides, and K. I. Bolotin, {\it The effect of intrinsic crumpling on the mechanics of free-standing graphene}, Nat. Comm. {\bf 6}, 8789 (2015).

\bibitem{Nicholl2017}  R. J. T. Nicholl, N. V. Lavrik, I. Vlassiouk, B. R. Srijanto, and K. I. Bolotin, {\it Hidden area and mechanical nonlinearities in freestanding graphene}, Phys. Rev. Lett. {\bf 118}, 266101  (2017).

\bibitem{Los2016} J. H. Los, A. Fasolino, and M. I. Katsnelson, {\it Scaling behavior and strain dependence of in--plane elastic properties of graphene}, Phys. Rev. Lett. {\bf 116}, 015901 (2016).

\bibitem {Gornyi:2015a} I. V. Gornyi, V. Y. Kachorovskii, and A. D.
    Mirlin, {\it Rippling and crumpling in disordered free-standing graphene}, Phys. Rev. B {\bf 92}, 155428 (2015).

\bibitem {Doussal2018} P. Le Doussal and L. Radzihovsky, {\it Anomalous elasticity, fluctuations and disorder in elastic membranes}, Ann. Phys.
    (N.Y.) {\bf 392}, 340 (2018).

\bibitem{Gornyi2017} I. V. Gornyi, V. Yu. Kachorovskii, and A. D. Mirlin, {\it Anomalous Hooke's law in disordered graphene}, 
    2D Mater. {\bf 4}, 011003 (2017).


\bibitem {Burmistrov2018b} I. S. Burmistrov, I. V. Gornyi, V. Y.
    Kachorovskii, M. I. Katsnelson, J. H. Los, and A. D. Mirlin, {\it Stress-controlled Poisson ratio of a crystalline membrane: Application to graphene}, Phys.
    Rev. B {\bf 97}, 125402 (2018).

\bibitem {Burmistrov2018a} I. S. Burmistrov, V. Y. Kachorovskii, I. V.
    Gornyi, and A. D. Mirlin, {\it Differential Poisson's ratio of a crystalline two-dimensional membrane}, Ann. Phys. (N.Y.) {\bf 396},  119 (2018).

\bibitem {Burmistrov2019a} D. R. Saykin, I. V. Gornyi, V. Y.
    Kachorovskii, and I. S. Burmistrov,  {\it Absolute Poisson's ratio and the bending rigidity exponent of a crystalline two-dimensional membrane}, arXiv:2002.04554.

\bibitem {Doussal1992} P. Le Doussal and L. Radzihovsky, {\it Self-consistent theory of polymerized membranes}, Phys. Rev. Lett.
    {\bf  69}, 1209 (1992).

\bibitem {Gazit2009} D. Gazit, {\it Structure of physical crystalline membranes within the self-consistent screening approximation}, Phys. Rev. E {\bf 80}, 041117 (2009).

\bibitem {Kownacki2009} J.-P. Kownacki and D. Mouhanna, {\it Crumpling transition and flat phase of polymerized phantom membranes}, Phys. Rev. E
    {\bf 79}, 040101 (2009).

\bibitem {Braghin2010} F. L. Braghin and N. Hasselmann, {\it Thermal fluctuations of free-standing graphene}, 
    Phys. Rev. B {\bf 82}, 035407 (2010).

\bibitem {Coquand:2018}  O. Coquand, K. Essafi, J.-P. Kownacki, and D. Mouhanna, {\it Glassy phase in quenched disordered crystalline membranes}, 
    Phys. Rev. E {\bf 97}, 030102(R) (2018).

\bibitem{Yeo2018} S. Yeo, J. Han, S. Bae, D. Su Lee, {\it Coherence in defect evolution data for the ion beam irradiated graphene}, 
    Scientific Reports {\bf  8}, 13973 (2018).

\bibitem{Daukiya2016}  H. Li, L. Daukiya, S. Haldar, A. Lindblad, B. Sanyal, O. Eriksson, D. Aubel, S. Hajjar-Garreau, L. Simon, and K. Leifer, {\it Site-selective local fluorination of graphene induced by focused ion beam irradiation}, Scientific Reports, {\bf  6}, 19719 (2016).

\bibitem{Yllanes2017}  D. Yllanes, S. S. Bhabesh, D. R. Nelson, and M. J. Bowick, {\it Thermal crumpling of perforated two-dimensional sheets}, Nat. Comm. {\bf 8},  1381 (2017).

\bibitem {Radzihovsky_1992} L. Radzihovsky and P. Le Doussal, {\it Crumpled glass phase of randomly polymerized membranes in the large d limit}, J. Phys. I {\bf 2}, 599 (1992).

\bibitem{Lopez2015} G. Lopez-Polin, C. Gomez-Navarro, V. Parente, F. Guinea, M. I. Katsnelson, F. Perez-Murano, and J. Gomez-Herrero, {\it Increasing the elastic modulus of graphene by controlled defect creation}, Nat. Phys. {\bf 11}, 26 (2015).

 \bibitem{Lopez2017} G. Lopez- Polin, M. Jaafar, F. Guinea,
    R. Roldan, C. Gomez- Navarro, and J. Gomez-Herrero, {\it The influence of strain on the elastic constants of graphene}, Carbon {\bf 124}, 42, (2017).

\bibitem{Zakharchenko2009} K. V. Zakharchenko, M. I. Katsnelson, and A.
    Fasolino, {\it Finite temperature lattice properties of graphene beyond the quasiharmonic approximation}, Phys. Rev. Lett. 102, 046808 (2009).
    
\bibitem{Bao2009} W. Bao, F. Miao, Z.
    Chen, H. Zhang, W. Jang, C. Dames, and C. N. Lau, {\it Controlled ripple texturing of suspended graphene and ultrathin graphite membranes}, Nat. Nanotech.
     {\bf 4}, 562 (2009).
     
   \bibitem{Yoom2011} D. Yoon, Y.-W. Son, and H. Cheong, {\it Negative thermal expansion coefficient of graphene measured by Raman spectroscopy}, Nano Lett. {\bf
    11}, 3227 (2011).
    
    \bibitem{Andres2012} P. L. de Andres, F. Guinea, and M. I. Katsnelson, {\it Bending modes, anharmonic effects, and thermal expansion coefficient in single-layer and multilayer graphene}, Phys. Rev. B {\bf 86}, 144103 (2012).
    
    \bibitem{Silva2014} A. L. C. da Silva, Ladir C\^andido, J.
    N. Teixeira Rabelo, G.-Q. Hai, and F. M. Peeters, {\it Anharmonic effects on thermodynamic properties of a graphene monolayer}, Europhys. Lett. (EPL),
    {\bf 107}, 56004 (2014).
    
\bibitem{Michel2015} K. H. Michel, S. Costamagna, and F. M.
    Peeters, {\it Theory of anharmonic phonons in two-dimensional crystals}, Phys. Rev. B {\bf 91}, 134302 (2015).
    
\bibitem{Burmistrov2016} I. S. Burmistrov, I. V. Gornyi, V. Y. Kachorovskii, M. I. Katsnelson, A. D. Mirlin, {\it Quantum elasticity of graphene: Thermal expansion coefficient and specific heat}, Phys. Rev. B {\bf 94}, 195430 (2016).

\bibitem{Bolotin2008} K. I. Bolotin, K. J. Sikes, J. Hone, H. L. Stormer, and P. Kim, {\it Temperature-dependent transport in suspended graphene}, Phys. Rev. Lett. {\bf 101}, 096802 (2008).

\bibitem{Castro2010} E. V. Castro, H. Ochoa, M. I. Katsnelson, R. V. Gorbachev, D. C. Elias, K. S. Novoselov, A. K. Geim, and F. Guinea, {\it Limits on charge carrier mobility in suspended graphene due to flexural phonons}, Phys. Rev. Lett.{\bf 105}, 266601
    (2010).
    
\bibitem{Gornyi2012} I. V. Gornyi, V. Yu. Kachorovskii, and A. D. Mirlin, {\it Conductivity of suspended graphene at the Dirac point}, Phys. Rev. B {\bf 86}, 165413  (2012).

\bibitem{Future} Will be published elsewhere.

\bibitem {See Supplemental Material} See Supplemental Material.

\bibitem{footnote} We note that analysis of the structure of the higher order in
$1/d_c$ diagrams suggests that the corrections to the position of the
unstable fixed point are $O(1/d_c)$ and, thus, are negligible at $d_c\gg
1$.

\bibitem{Comment-Mohana} Recently,
the unstable fixed point similar to $f_*$ has been found for a disordered
membrane of $D=4-\epsilon$ dimension within the second order expansion in
$\epsilon$ and  for a 2D disordered membrane within analytically uncontrolled  NPRG approach \cite{Coquand:2018}.


\bibitem{Nishimori} H. Nishimori, \textit{Statistical Physics of Spin Glasses and
Information Processing. An Introduction}, Clarendon Press, Oxford (2001).

\bibitem{LeDoussal1988} P. Le Doussal, A. Brooks Harris, {\it Location of the Ising Spin-Glass Multicritical Point on Nishimori's Line}, Phys. Rev. Lett. {\bf 61}, 625 (1988).

\bibitem{LD&R1993} P. Le Doussal and  L. Radzihovsky, {\it Flat glassy phases and wrinkling of polymerized membranes with long-range disorder}, Phys. Rev. B {\bf 48}, 3548 (1993).



\end{thebibliography}

\begin{thebibliography}{100}
\newcounter{Sbib}

\stepcounter{Sbib}
\bibitem[S\theSbib]{AOP2020} D. R. Saykin, I. V. Gornyi, V. Y. Kachorovskii, and I. S.
Burmistrov, arXiv:2002.04554.

\bibitem[S\theSbib]{Morse} D. C. Morse and T. C. Lubensky, Phys. Rev. A {\bf 46}, 1751 (1992).

\bibitem[S\theSbib]{Gornyi}I. V. Gornyi, V. Y. Kachorovskii, and A. D. Mirlin, Phys.
Rev. B {\bf 92}, 155428 (2015).


\end{thebibliography}

\newpage

\onecolumngrid
\begin{center}
	\large
	\bf 
	ONLINE SUPPORTING INFORMATION\\[6pt]
	Disorder-induced rippled phases and multicriticality in a free-standing graphene
\end{center}
\begin{center}
	In this Supplementary Material we present derivation of the RG equations (4) of the main text.
\end{center}

\section{Self-energy correction}

The interaction between flexural phonons modifies the Green's function. The exact Green's function can be written as follows
(in the replica limit $N\to 0$):
\begin{equation}
\hat{\bm{\mathcal{G}}}(k)= T \Bigl [\varkappa (1-f \hat J) k^4 -\hat \Sigma(k)\Bigr ]^{-1} .
\end{equation}

As well-known, before constructing the perturbation theory in the interaction between flexural phonons it is important to take into account screening of this interaction by the flexural phonon themselves.
This screening (see Fig. 2b of the main text) is determined by the bare polarization operator
\begin{gather}
\hat{\Pi}_{ab}(q) = \frac{d_c}{3T}
\int_k \frac{[\bm{k \times q}]^4}{q^4}
\hat{\mathcal{G}}_{ab}(|\bm{k-q}|)\hat{\mathcal{G}}_{ab}(k) = \Bigl(1+2f+f^2 \hat J\Bigr ) \frac{d_c T}{16\pi \varkappa^2 q^2} .
\end{gather}
Here we introduced for a brevity the following shorthand notation: $\int_k \equiv \int d^2\bm{k}/(2\pi)^2$. Summation of the geometric series shown in Fig. 2b of the main text yields the screened interaction
\begin{gather}
\hat N(q) = \frac{\hat Y_q}{2}= \frac{Y_0/2}{1+
3 Y_0 \hat{\Pi}(q) /2}
= \frac{Y_0}{2}\frac{q^2}{q^2+\tilde{q}_*^2}
\left ( 1 - \frac{f^2 \hat J}{1+2f} +  \frac{q^2}{q^2+\tilde{q}_*^2}\frac{f^2 \hat J}{1+2f} \right )  .
\end{gather}
We mention that the screened interaction at small momenta,
$$q\ll \tilde{q}_*=\sqrt{1+2 f_0}/L_* $$ 
becomes independent of the Young modulus $Y_0$ and proportional to $1/d_c$.

\subsection{Contribution of the first order in $1/d_c$}

The self-energy correction of the first order in $1/d_c$ is given by the diagram in Fig.  2a of the main text. It can be written as:
\begin{align}
	\begin{aligned}
		\hat\Sigma^{(1)}_{ab}(k) 
		&= - 2 \int_q \frac{[\bm{k \times q}]^4}{q^4}
		\hat N_{ab}(q) \hat{\mathcal{G}}_{ab}(|\bm{k-q}|)	\\
		&= - \frac{2}{d_c} \varkappa k^4
		 \Biggl [  \frac{1+3f +f^2 -f^3 \hat{J}}{(1+2f)^2} L_0\left(\frac{k}{\tilde{q}_*}\right )
		+ \frac{f^2(1+f\hat J)}{(1+2f)^2}  L_1\left(\frac{k}{\tilde{q}_*}\right ) \Biggr ]_{ab}
	\end{aligned}
	 \label{eq:Sigma:1loop}
\end{align}
where
\begin{gather}
L_m(K)= \int_0^\infty \frac{dq}{q} \frac{q^{2m}\min\bigl \{{q^4}/{K^4},1\bigr \}}{(1+q^2)^{m+1}}.
\end{gather}
For $m=0,1$ they are given explicitly as follows
\begin{equation}
L_0(K)  =
-\ln K + \frac{1}{2}\ln(1+K^2) + \frac{K^2-\ln(1+K^2)}{2K^4} ,
\quad
L_1(K)  =\frac{K^2-\ln(1+K^2)}{K^4} .
\end{equation}
In the limit $k/\tilde{q}_*\ll 1$ and $N\to 0$, we find
\begin{equation}
\hat \Sigma^{(1)}(k) = - \frac{2}{d_c}  \varkappa k^4
\bigl (\alpha_1 \ln \frac{q_*}{k}  - f \gamma_1\hat J
\ln \frac{q_*^\prime}{k}
\bigr )
  ,
\label{eq:sigma:1}
\end{equation}
where $q_* = \tilde{q}_* \exp[1/4+\gamma_1/(2\alpha_1)]$ and $q_*^\prime = \tilde{q}_* \exp(-1/4)$ and
\begin{equation}
\alpha_1  = \frac{1+3f +f^2}{(1+2f)^2} , \quad
\gamma_1 = \frac{f^2}{(1+2f)^2} .
\end{equation}

\subsection{Contribution of the second order in $1/d_c$}

In this subsection we present results for the contribution of the second order in $1/d_c$ to the self-energy (see diagrams in Fig. 3 of the main text)

\subsubsection{Diagram Fig. 3a}

The corresponding contribution to the self-energy has the following form
\begin{gather}
\hat \Sigma^{(2,a)}_{ab}(k) = - \frac{2}{T} \int_q  \Bigl [\hat{\mathcal{G}}(|\bm{k-q}|)\Sigma^{(1)}(|\bm{k-q}|)\hat{\mathcal{G}}(|\bm{k-q}|)\Bigr ]_{ab}
\frac{[\bm{k \times q}]^4}{q^4}
\hat N_{ab}(q)
 .
\label{eq:Sigma:2a:t1}
\end{gather}
Computing the integrals over momentum $q$ in the same way as in Ref. \cite{AOP2020}, we obtain for $k/\tilde{q}_*\ll 1$ and $N\to 0$:
\begin{gather}
\hat \Sigma^{(2,a)} (k)
= \frac{2}{d_c^2} \varkappa k^4 \Biggl\{ \alpha^{(a)}
 \Bigl [\ln^2 \frac{q_*}{k}
+\frac{1}{2} \ln \frac{q_*}{k}\Bigr ] -
f \gamma^{(a)}\hat J
 \Bigl [\ln^2 \frac{q_*^\prime}{k}
+\frac{1}{2} \ln \frac{q_*^\prime}{k}\Bigr ]
+\tilde{\alpha}^{(a)} \ln \frac{q_*}{k}- f \tilde{\gamma}^{(a)}\hat J \ln \frac{q_*^\prime}{k}
 \Biggr \} ,
%
\label{eq:Sigma:2a}
\end{gather}
where
\begin{align}
\alpha^{(a)} & = \frac{1+7f+16f^2+12 f^3+f^4}{(1+2f)^4} , \quad
\gamma^{(a)}   =\frac{f^2(2+6 f+f^2)}{(1+2f)^4} ,
\notag\\
\tilde{\alpha}^{(a)} & = -\frac{c \gamma_1}{\beta_1} \beta^{(2,a)}+ \frac{f^2(1+5f+5f^2)}{(1+2f)^4} ,
\quad \tilde{\gamma}^{(a)}   = {\gamma}^{(2,a)}
+\frac{3 f^4}{(1+2f)^4}
.
\end{align}

\subsubsection{Diagram Fig. 3b}

The diagram can be considered as the first order correction to the self-energy in which the interaction line is changed due to correction to the polarization operator:
\begin{gather}
\hat\Sigma^{(2,b)}_{ab}(k) =\! \frac{6}{T}\!\! \int_q \!\frac{[\bm{k \times q}]^4}{q^4}
\Bigl [ \hat N(q) \delta \hat \Pi(q)\hat N(q) \Bigr ]_{ab} \!\hat{\mathcal{G}}_{ab}(|\bm{k-q}|)
,
\notag\\
 \delta \hat \Pi_{cd}(q) = \frac{2d_c}{3T^2}
\int_k \frac{[\bm{k \times q}]^4}{q^4}
\hat{\mathcal{G}}_{cd}(|\bm{k-q}|)
\Bigl [ \hat{\mathcal{G}}(k) \hat\Sigma^{(1)}(k)\hat{\mathcal{G}}(k) \Bigr]_{cd}
 .
 \label{eq:Sigma2b:start}
\end{gather}
The correction to the polarization operator can be computed as follows:
\begin{gather}
\delta \hat \Pi(q) =- \frac{T \delta \hat \pi(q/\tilde{q}_*) }{4\pi \varkappa^2q^2}
 , \quad
 \delta \hat \pi(q) = \sum_{j=0,1}
 \delta\hat \pi_j \tilde{L}_j(q)
 ,
 \label{eq:deltaPi}
 \end{gather}
 where
 \begin{equation}
 \delta\hat \pi_0  =\frac{1+6f+10f^2+2f^3+f^2(2+6f+f^2)\hat J}{(1+2f)^2}
 ,
\quad
\delta\hat \pi_1   = \frac{f^2(1+4f+3f^2\hat J)}{(1+2f)^2}
 ,
\end{equation}
and
\begin{equation}
\tilde{L}_m(q)=\int_0^1 dk k L_m(k q)+
\int_1^\infty \frac{dk}{k^3}  L_m(k q) .
\end{equation}
The functions $\tilde L_0$ and $\tilde L_1$ can be computed exactly as follows
\begin{gather}
\tilde L_0(q) =
\frac{(1+q^2)}{6q^2} \Bigl[\frac{(1+q^2)^2}{q^2} \ln(1+q^2)
-1\Bigr ] - \frac{q^2+3}{3} \ln q
,
\notag \\
\tilde{L}_1(q) = \frac{q^2}{3} \ln q
- \frac{(q^2-2)}{6 q^2}\Bigl[ \frac{(1+q^2)^2}{q^2}  \ln(1+q^2)
-1\Bigr ] .
\end{gather}
Substituting the expression \eqref{eq:deltaPi} for the correction to the polarization operator into Eq. \eqref{eq:Sigma2b:start},  we obtain in the limits $k/\tilde{q}_*\ll 1$ and $N\to 0$:
\begin{gather}
\hat\Sigma^{(2,b)}(k) =
- \frac{4}{d_c^2}  \varkappa k^4
\Biggl \{\alpha^{(b)}\Bigl[ \ln^2\frac{q_*}{k}+\frac{1}{2} \ln \frac{q_*}{k}\Bigr ]
-f \gamma^{(b)} \hat J\Bigl[ \ln^2\frac{q_*^\prime}{k}+\frac{1}{2} \ln \frac{q_*^\prime}{k}\Bigr ]
+\tilde{\alpha}^{(b)} \ln \frac{q_*}{k} -f \tilde{\gamma}^{(b)} \hat J \ln \frac{q_*^\prime}{k}
\Biggr \}
\label{eq:Sigma:2b}
\end{gather}
where
\begin{align}
\alpha^{(b)} & =
\frac{1 + 9 f + 30 f^2 + 42 f^3 + 19 f^4 + 2 f^5}{(1+2f)^5} ,
\quad
 \gamma^{(b)}  = \frac{f^3(2 + 7 f + 2 f^2)}{(1+2f)^5} ,
 \notag \\
 \tilde\alpha^{(b)} & = -\frac{c \gamma_1}{\beta}
 \beta^{(b)} +
 \frac{f^2(1 +7 f + 15 f^2 + 6 f^3)}{(1+2f)^5},  \quad
 \tilde\gamma^{(b)}  =  \gamma^{(b)}-c \frac{f^4(1-2f)}{(1+2f)^5} .
\end{align}

\subsubsection{Diagram Fig. 3c}

The correction to the self-energy shown in Fig. 3c of the main text can be written as follows
\begin{gather}
\hat \Sigma^{(2,c)}_{ab}(k) = \frac{4}{T}
\int_{q,Q} \frac{[\bm{k \times q}]^2}{q^2}
 \frac{[\bm{k \times Q}]^2}{q^2} \frac{[\bm{(k-q) \times Q}]^2}{Q^2}
 \frac{[\bm{(k-Q) \times q}]^2}{q^2}
 \hat{\mathcal{G}}_{ac}(|\bm{k-q}|)\hat{\mathcal{G}}_{cd}(|\bm{k-Q}|)
 \notag \\
 \times \hat{\mathcal{G}}_{db}(|\bm{k-q-Q}|)\hat N_{ad}(q)\hat N_{cb}(Q)  .
 \end{gather}
Taking the integrals over momenta in the same way as in Ref. \cite{AOP2020}, we find
in the limits $k/\tilde{q}_*\ll 1$ and $N\to 0$:
\begin{gather}
\hat \Sigma^{(2,c)}(k) =\frac{7}{3 d_c^2}\varkappa k^4
\bigl (\alpha^{(c)}-f \gamma ^{(c)}\hat J \bigr ) \ln \frac{q_*}{k} ,
\label{eq:Sigma:2c}
\end{gather}
where
\begin{equation}
\alpha^{(c)}  = \frac{1+7f+17f^2+16f^3+5f^4}{(1+2f)^4} , \quad
\gamma ^{(c)}   =\frac{f^2(1+4f+f^2)}{(1+2f)^4} .
\end{equation}

\subsubsection{Diagram Fig. 3d}

The correction to the self-energy shown in Fig. 3d of the main text can be written as follows
\begin{gather}
\hat \Sigma^{(2,d)}_{ab}(k) = - \frac{4 d_c}{T^2}
\int_{Q,p,p^\prime}
\frac{[\bm{p \times Q}]^2}{Q^2} \frac{[\bm{p^\prime \times Q}]^2}{Q^2}
\frac{[\bm{p \times p^\prime}]^2}{|\bm{p-p^\prime}|^2}
\frac{[\bm{k \times (p-p^\prime)}]^4}{|\bm{p-p^\prime}|^4}
 \frac{[\bm{(p-Q) \times (p^\prime-Q)}]^2}{|\bm{p-p^\prime}|^2}
 \hat N_{lm}(Q)
 \notag \\
 \times
 \hat N_{ac}(|\bm{p-p^\prime}|)\hat N_{bd}(|\bm{p-p^\prime}|)
 \hat{\mathcal{G}}_{cl}(p) \hat{\mathcal{G}}_{dl}(|\bm{p-Q}|)
  \hat{\mathcal{G}}_{cm}(p^\prime)\hat{\mathcal{G}}_{dm}(|\bm{p^\prime-Q}|)\hat{\mathcal{G}}_{ab}(|\bm{k-p+p^\prime}|)
 .
 \end{gather}
 Integration over momenta can be performed in the same way as in Ref. \cite{AOP2020}. Then for $k/\tilde{q}_*\ll 1$ and $N\to 0$ we retrieve:
\begin{gather}
\hat \Sigma^{(2,d)}(k) =- \frac{2}{d_c^2}\varkappa k^4
\bigl (\alpha^{(d)}-f \gamma ^{(d)}\hat J \bigr ) \ln \frac{q_*}{k} ,
\label{eq:Sigma:2d}
\end{gather}
where
\begin{equation}
\alpha^{(d)}  = \frac{1+11f+49f^2+111f^3+130f^4+72f^5+16 f^6}{(1+2f)^6} , \quad
\gamma ^{(d)}   =\frac{2f^4}{(1+2f)^5} .
\end{equation}

\subsubsection{Diagram Fig. 3e}

The correction to the self-energy shown in Fig. 3e of the main text is as follows
\begin{gather}
\hat \Sigma^{(2,e)}_{ab}(k) = - \frac{8 d_c}{T^2}
\int_{q,p,Q}
 \frac{[\bm{k \times q}]^2}{q^2}
 \frac{[\bm{k \times Q}]^2}{Q^2}
  \frac{[\bm{p \times q}]^2}{q^2}
  \frac{[\bm{(k-q) \times (Q-q)}]^2}{|\bm{q-Q}|^2}
 \frac{[\bm{(p-Q) \times (p-q)}]^2}{|\bm{q-Q}|^2}
 \notag \\
 \times
 \frac{[\bm{p \times Q}]^2}{Q^2}
 \hat N_{ac}(q)\hat N_{bd}(Q)
\hat N_{lm}(|\bm{q-Q}|)
\hat{\mathcal{G}}_{cd}(p)
\hat{\mathcal{G}}_{am}(|\bm{k-q}|)\hat{\mathcal{G}}_{mb}(|\bm{k-Q}|)
\hat{\mathcal{G}}_{cl}(|\bm{p-q}|)
\hat{\mathcal{G}}_{dl}(|\bm{p-Q}|)  .
 \end{gather}
Integration over momenta can be performed in the same way as in Ref. \cite{AOP2020}. Then we obtain for $k/\tilde{q}_*\ll 1$ and $N\to 0$:
\begin{gather}
\hat \Sigma^{(2,e)}(k) =- \frac{58}{27 d_c^2} \varkappa k^4
\bigl (\alpha^{(e)}-f \gamma ^{(e)}\hat J \bigr ) \ln \frac{q_*}{k} ,
\label{eq:Sigma:2e}
\end{gather}
where
\begin{equation}
\alpha^{(e)}  = \frac{1+11f+49f^2+111f^3+129f^4+67f^5+10f^6}{(1+2f)^6} , \quad
\gamma ^{(e)}   =\frac{2f^4(2+5f+2f^2)}{(1+2f)^6} .
\end{equation}

\subsubsection{Diagram Fig. 3f}

The correction to the self-energy shown in Fig. 3f of the main text is given by the following explicit expression:
\begin{gather}
\hat \Sigma^{(2,f)}_{ab}(k) = \frac{8 d^2_c}{T^3}
\int_{p,p^\prime,q,Q}
 \frac{[\bm{k \times q}]^4}{q^4}
   \frac{[\bm{p \times q}]^2}{q^2} \frac{[\bm{p \times Q}]^2}{Q^2}
 \frac{[\bm{(p-Q) \times (q-Q)}]^2}{|\bm{q-Q}|^2}
 \frac{[\bm{(p^\prime-Q) \times (q-Q)}]^2}{|\bm{q-Q}|^2}
 \notag \\
 \times
 \frac{[\bm{p^\prime \times Q}]^2}{Q^2}
 \frac{[\bm{p^\prime \times q}]^2}{q^2}
\hat N_{ac}(q) \hat N_{bd}(q)
\hat N_{st}(|\bm{q-Q}|)
\hat N_{lm}(Q)
 \hat{\mathcal{G}}_{cl}(p) \hat{\mathcal{G}}_{cs}(|\bm{p-Q}|)
 \hat{\mathcal{G}}_{ls}(|\bm{p-q}|)
 \hat{\mathcal{G}}_{dm}(p^\prime)
  \notag \\
  \times
 \hat{\mathcal{G}}_{dt}(|\bm{p^\prime-Q}|) \hat{\mathcal{G}}_{mt}(|\bm{p^\prime-q}|)
  \hat{\mathcal{G}}_{ab}(|\bm{k-q}|).
 \end{gather}
 Integrating over momenta in the same way as in Ref. \cite{AOP2020}, and taking the limits $k/\tilde{q}_*\ll 1$ and $N\to 0$, we obtain:
\begin{gather}
\hat \Sigma^{(2,f)}_{ab}(k) =
\frac{3+68\zeta(3)}{27 d_c^2} \varkappa k^4
\bigl (\alpha^{(f)}-f \gamma ^{(f)}\hat J \bigr )
\ln \frac{q_*}{k} ,
\label{eq:Sigma:2f}
\end{gather}
where
\begin{equation}
\alpha^{(f)}  = \frac{1+11f+49f^2+111f^3+129f^4+64f^5+9f^6}{(1+2f)^6} , \quad
\gamma ^{(f)}   =\frac{f^4(1+6f+3f^2)}{(1+2f)^6} .
\end{equation}

\subsubsection{Contribution of the second order in $1/d_c$}

All in all, the six diagrams in Fig. 3 of the main text yield the following contribution to the self-energy in the second order in $1/d_c$:
\begin{equation}
\hat\Sigma^{(2)}(k) =  -
 \frac{\alpha_2+2\tilde{\alpha}_2}{d_c^2}  \ln \frac{q_*}{k}
- \frac{2 \alpha_2^\prime}{d_c^2}
\ln^2 \frac{q_*}{k}
 + f \hat J \Biggl [
 \frac{\gamma_2+2\tilde{\gamma}_2}{d_c^2} \ln \frac{q_*^\prime}{k}
+ \frac{2 \gamma_2^\prime}{d_c^2}
\ln^2 \frac{q_*^\prime}{k}\Biggr ]
,
\label{eq:SelfEnergy:final}
\end{equation}
where
\begin{equation}
\alpha_2^\prime  =2\alpha^{(b)}-\alpha^{(a)}, \quad
\tilde{\alpha}_2  =2 \tilde\alpha^{(b)}-\tilde\alpha^{(a)},
\quad
\gamma_2^\prime  = 2\gamma^{(b)}-\gamma^{(a)},
\quad
 \tilde{\gamma}_2=2 \tilde\gamma^{(b)}-\tilde\gamma^{(a)} .
\end{equation}
and
\begin{align}
\alpha_2 & =
\alpha_2^\prime
-\frac{7}{3} \alpha^{(c)}
+2 \alpha^{(d)}+\frac{58}{27} \alpha^{(e)}
-\frac{3+68\zeta(3)}{27} \alpha^{(f)} ,
\notag \\
\gamma_2 &= \gamma_2^\prime -\frac{7}{3} \gamma^{(c)}
+2 \gamma^{(d)}+\frac{58}{27} \gamma^{(e)}
-\frac{3+68\zeta(3)}{27} \gamma^{(f)} .
\end{align}

\section{RG equations}

\subsection{First order in $1/d_c$}

 The effect of the first order correction \eqref{eq:sigma:1} to the self-energy can be interpreted as $1/d_c$ corrections to $\varkappa$ and $f$:
\begin{equation}
\varkappa(k)  = \varkappa \left [1+\frac{2 \alpha_1}{d_c} \ln \frac{q_*}{k} \right ] ,\quad \varkappa(k) f(k)  = \varkappa f \left [1+\frac{2 \gamma_1}{d_c} \ln \frac{q_*^\prime}{k} \right ], \quad
f(k)  =  f \left [1+\frac{2 (\gamma_1-\alpha_1)}{d_c} \ln \frac{q_*^{\prime\prime}}{k} \right ]
 ,
\label{eq:1loop:PR}
\end{equation}
where $q_*^{\prime\prime}=q_* e^{1/4+c\gamma_1/(\alpha_1-\gamma_1)}$.
We mention that the perturbative results \eqref{eq:1loop:PR} suggest that the bare parameters $\varkappa$ and $f$ coincide with the renormalized parameters at the scales $q_*$ and $q_*^{\prime\prime}$, respectively, i.e.
$\varkappa\equiv \varkappa(q_*)$ and  $f\equiv f(q_*^{\prime\prime})$. Also we note that the following relation holds $\varkappa(q_*)f(q_*^{\prime\prime}) = \varkappa(q_*^{\prime}) f(q_*^\prime)$.

The perturbative corrections \eqref{eq:1loop:PR} can be recast in the form of the RG equations \cite{Morse,Gornyi}:
\begin{equation}
-\frac{d\ln \varkappa}{d\ln k} =  \frac{\eta_\varkappa^{(1)}}{d_c},\qquad
-\frac{d\ln f}{d\ln k} = \frac{\beta^{(1)}}{d_c} ,\qquad\qquad \eta_\varkappa^{(1)} = 2\alpha_1, \qquad \beta^{(1)} =2(\gamma_1-\alpha_1).
\label{eq:1loop:RG}
\end{equation}
 Since the momentum scale are arranged as $q_*^{\prime}<q_*<q_*^{\prime\prime}$, strictly speaking, the RG equations \eqref{eq:1loop:RG} are valid for $k<q_*^\prime$.

\subsection{Second order in $1/d_c$}

The results \eqref{eq:sigma:1} and \eqref{eq:SelfEnergy:final} for the self-energy allows us to write the following perturbative expansions for bending rigidity $\varkappa$ and the parameter $f$:
\begin{align}
\frac{\varkappa(k)}{\varkappa(q_*)} = &
1+\frac{2\alpha_1}{d_c}  \ln \frac{q_*}{k}
+ \frac{1}{d_c^2}\Bigl(\alpha_2+4 f \gamma_1 \frac{d\alpha_1}{df}
\ln \frac{q_*^\prime}{q_*}
\Bigr ) \ln \frac{q_*}{k}
+
\frac{2}{d_c^2} \Bigl ( \alpha_1^2+(\gamma_1-\alpha_1)f \frac{d\alpha_1}{df}\Bigr )
\ln^2 \frac{q_*}{k}
\label{eq:kappa:1}
\\
\frac{f(k)}{f(q_*^{\prime\prime})}   =  &
1+\left (\frac{2(\gamma_1-\alpha_1)}{d_c}  +
\frac{(\gamma_2-\alpha_2)}{d_c^2} \right ) \ln \frac{q_*^{\prime\prime}}{k}
+
\frac{2(\gamma_1-\alpha_1)}{d_c^2} \frac{d [f(\gamma_1-\alpha_1)]}{df}
\ln^2 \frac{q_*^{\prime\prime}}{k}
 .
\label{eq:kappa:2}
\end{align}
We emphasize that $f$ in the right hand side (r.h.s.) of Eqs. \eqref{eq:kappa:1} and \eqref{eq:kappa:2} is defined at the momentum scale $q_*^{\prime\prime}$. Also we note that
in derivation of Eqs. \eqref{eq:kappa:1} and \eqref{eq:kappa:2}
we used the following non-trivial relations:
\begin{equation}
\alpha_2^\prime  = \alpha_1^2+(\gamma_1-\alpha_1)f \frac{d\alpha_1}{df} ,\quad
\gamma_2^\prime  =
\gamma_1^2+(\gamma_1-\alpha_1)f \frac{d\gamma_1}{df} .
\end{equation}
The perturbative expansion \eqref{eq:kappa:2} describes how the disorder parameter $f$ transforms under a change of the momentum scale from $q_*^{\prime\prime}$ to $k$. The form of Eq. \eqref{eq:kappa:1} is a bit unconventional since its r.h.s. involves $f$ at not at the momentum scale $q_*$ but at another momentum scale, $q_*^{\prime\prime}$. Therefore it is convenient to rewrite Eq. \eqref{eq:kappa:1} with $f$ defined at the momentum scale $q_*$ in its r.h.s.:
\begin{equation}
\frac{\varkappa(k)}{\varkappa(q_*)} =
1+\left (\frac{2\alpha_1}{d_c}
+ \frac{\alpha_2}{d_c^2} \right ) \ln \frac{q_*}{k}
+
\frac{2}{d_c^2} \Bigl ( \alpha_1^2+(\gamma_1-\alpha_1)f \frac{d\alpha_1}{df}\Bigr )
\ln^2 \frac{q_*}{k} .
\label{eq:kappa:10}
\end{equation}
We stress that $f$ in the r.h.s. of Eq. \eqref{eq:kappa:10} is defined at the momentum scale $q_*$.

The results \eqref{eq:kappa:2} and \eqref{eq:kappa:10} can be cast from perturbative solutions of the following RG equations for $\varkappa$ and $f$:
\begin{equation}
-\frac{d\ln \varkappa}{d\ln k} = \frac{\eta_\varkappa^{(1)}}{d_c}
 +
\frac{\eta_\varkappa^{(2)}}{d_c^2}  ,
\qquad
-\frac{d\ln f}{d\ln k}  = \frac{\beta^{(1)}}{d_c}+
\frac{\beta^{(2)}}{d_c^2} , \qquad \qquad
\eta_\varkappa^{(2)}=\alpha_2, \qquad\beta^{(2)}=\alpha_2-\gamma_2 .
\label{eq:final}
\end{equation}
RG equations \eqref{eq:final} equivalent to Eqs. (4) of the main text.

\end{document}